\documentclass[aps, prl, reprint, 10pt]{revtex4-1}
\usepackage{graphicx}
\usepackage{amsmath}

\begin{document}
\title{Local control of single atom magneto-crystalline anisotropy}
\author{B. Bryant}\thanks{B. Bryant and A. Spinelli contributed equally to this work.}
\author{A. Spinelli}\thanks{B. Bryant and A. Spinelli contributed equally to this work.}
\author{J. J. T. Wagenaar}
\author{M. Gerrits}
\author{A. F. Otte}\thanks{a.f.otte@tudelft.nl}
\affiliation{Delft University of Technology, Kavli Institute of Nanoscience, Department of Quantum Nanoscience, Lorentzweg 1, 2628 CJ Delft, The Netherlands}

\begin{abstract}
Individual Fe atoms on a Cu$_2$N/Cu(100) surface exhibit strong magnetic anisotropy due to the crystal field. Using atom manipulation in a low-temperature STM we demonstrate that the anisotropy of one Fe atom is significantly influenced by local strain due to a second Fe atom placed nearby. Depending on the relative positions of the two atoms on the Cu$_2$N lattice we can controllably enhance or reduce the uniaxial anisotropy. We present a model that explains the observed behavior qualitatively in terms of first principles.
\end{abstract}
\maketitle

\begin{figure*}[htbp]
\begin{centering}
\includegraphics{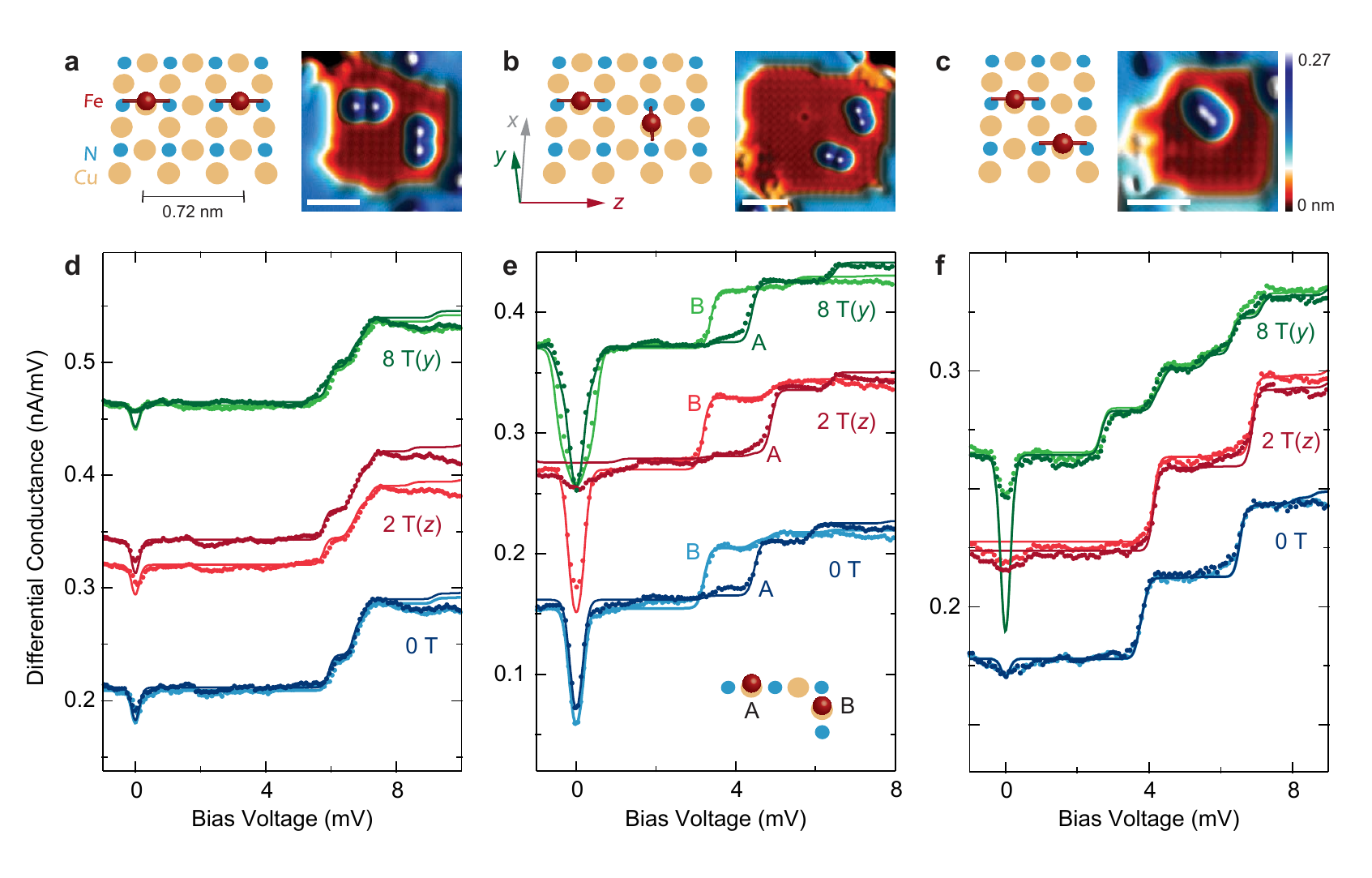}
\caption{(color online). (a--c) Diagrams and STM topographic images for: (a) $\left\{2,0\right\}$ dimer, (b) $\left\{\frac{3}{2},\frac{1}{2}\right\}$ dimer and (c) $\left\{1,1\right\}$ dimer. Magnetic easy axes for Fe atoms are indicated. Topographic scale bars correspond to 2 nm. (d--f) Measured IETS spectra (dots) and corresponding simulated spectra (lines) on all three types of dimers, for zero magnetic field and magnetic fields applied in $z$ and $y$ directions. In (e) atoms A and B are distinguished: in (d) and (f) the atoms of the dimer are essentially identical, but both sets of spectra are presented for comparative purposes. The discrepancy between measured and simulated spectra near zero bias in (f) for 8~T($y$) could be accounted for by a $\sim2^{\circ}$ misalignment of the magnetic field to the $y$~axis.}
\label{Figure1}
\end{centering}
\end{figure*}

The macroscopic magnetic behavior of materials is ultimately dictated by the manner in which individual magnetic atoms interact with their environment. Recent experimental advances such as the ability to probe and manipulate individual magnetic atoms using scanning tunneling microscopy (STM) \cite{Heinrich04,Lee04,Kitchen06} make it possible to investigate these interactions in detail. In general, two competing processes can be identified, which both influence the preferred orientation of atomic spins. On one hand, magnetic anisotropy due to the local crystal field favors certain axes for magnetization over others \cite{Gambardella03}. This effect plays an important role in particular in low-symmetry environments such as covalent structures \cite{Hirjibehedin07,Tsukahara09} and atomic clusters \cite{Balashov09}. On the other hand, neighboring spins can be subject to spin coupling due to \textit{e.g.} superexchange \cite{Hirjibehedin06,Chen08} or RKKY interaction \cite{Zhou10,Neel11,Khajetoorians12} leading to either ferromagnetic or antiferromagnetic alignment of the spins.

An interesting situation arises when spin coupling and magnetic anisotropy energies are comparable \cite{Otte09, OttePhD, Loth12}. Recently it was shown that  placing only a few atoms in this type of configuration results in remarkably stable magnetic structures \cite{Loth12}. Understanding the physical mechanisms underlying this sudden emergence of magnetic stability is of great importance for the development of nanoscale data storage solutions.

In this Letter we demonstrate that the magnetic anisotropy of one magnetic atom placed in a covalent surface network is significantly influenced by local strain due to a second magnetic atom placed nearby. We present a family of Fe dimers built on Cu$_2$N using atom manipulation in a low-temperature STM. The three dimers, which all have comparable interatomic spacing, were specifically designed so that we can make a distinction between spin coupling and magnetic anisotropy: some atoms have their primary magnetization axis parallel to the dimer axis whereas for others the two are at an angle. Depending on the relative position and orientation of the magnetic easy axes of the atoms of the dimer, the uniaxial anisotropy parameter $D$ is either enhanced or reduced by values up to $20\%$.

Atomic structure and STM topographic images of each of the three types of dimers are shown in Figs.~1(a--c). We classify the dimers according to the number of unit cells separating the two atoms in each symmetry direction. For example, the dimer shown in Fig.~1b will be referred to as $\left\{\frac{3}{2},\frac{1}{2}\right\}$. Each Fe atom on Cu$_2$N has an easy axis for magnetization ($D<0$), which is oriented toward the neighboring N atoms \cite{Hirjibehedin07}. As such, the atoms in the $\left\{\frac{3}{2},\frac{1}{2}\right\}$ dimer have their easy axes oriented perpendicular to each other. The linear $\left\{2,0\right\}$ dimer (Fig.~1a) has both easy axes in-line. This structure is identical to the atomic arrangement described previously by Loth \textit{et al.} \cite{Loth12}. Finally, the $\left\{1,1\right\}$ dimer (Fig.~1c) has parallel easy axes as well, but here the two axes are offset with respect to each other.

We performed inelastic electron tunneling spectroscopy (IETS) measurements on each of the atoms in our Fe dimers (Figs.~1d--f). In the resulting differential conductance spectra, spin excitations appear as distinct steps at voltages corresponding to the excitation energies: these energies may vary with applied magnetic field. The measured spectra are markedly different from those found on isolated Fe atoms on Cu$_2$N \cite{Hirjibehedin07}. In the following we demonstrate how these differences can be accounted for in terms of spin coupling, and modifications to the local crystal field.

Observed excitation energies can be modeled using a Heisenberg Hamiltonian \cite{Rossier09}
\begin{equation}
\hat{\mathcal{H}}=\sum\limits_{i={\rm A,B}}\hat{\mathcal{H}}^{(i)}+J\hat{\bf S}^{\rm (A)}\cdot\hat{\bf S}^{\rm (B)},
\label{Heisenberg-Hamiltonian}
\end{equation}
which couples the spins ${\bf S}^{(i)}$ of atoms A and B in the dimer through a Heisenberg coupling parameter $J$. The single spin anisotropy Hamiltonian $\hat{\mathcal{H}}^{(i)}$ describes for each spin the magnetic anisotropy and the Zeeman effect due to an external magnetic field ${\bf B}$ ($\mu_{\rm B}$ being the Bohr magneton):
\begin{equation}
\begin{split}
\hat{\mathcal{H}}^{(i)}=D^{(i)}\hat{S}_{z}^{2(i)} &+ E^{(i)}\left(\hat{S}_{x}^{2(i)}-\hat{S}_{y}^{2(i)}\right) \\
&-\mu_{\rm B}\sum\limits_{\mu=x,y,z}g_{\mu}^{(i)}B_{\mu}\hat{S}_{\mu}^{(i)}.
\end{split}
\label{single-anisotropy-Hamiltonian}
\end{equation}

Here the anisotropy parameters $D^{(i)}$ and $E^{(i)}$ as well as the g-tensor $g_{\mu}^{(i)}$ follow from second order perturbation treatment of the spin-orbit coupling $\lambda {\bf L}^{(i)} \cdot {\bf S}^{(i)}$ \cite{Dai08}:
\begin{eqnarray}
D^{(i)} &=& -\frac{\lambda^2}{2}\left(2\Lambda_{zz}^{(i)}-\Lambda_{xx}^{(i)}-\Lambda_{yy}^{(i)}\right), \\
E^{(i)} &=& -\frac{\lambda^2}{2}\left(\Lambda_{xx}^{(i)}-\Lambda_{yy}^{(i)}\right), \\
g_{\mu}^{(i)} &=& 2\left(1-\lambda\Lambda_{\mu\mu}^{(i)}\right).
\label{coefficients}
\end{eqnarray}

In these expressions the parameters $\Lambda_{\mu\mu}^{(i)}$ represent the degree of unquenched orbital momentum along the $\mu$-direction (where $\mu=x,y,z$). These are defined as:
\begin{equation}
\Lambda_{\mu\mu}^{(i)}\equiv\sum\limits_{n}\frac{\left|\left\langle\psi_{0}\left|\hat{L}_{\mu}^{(i)}\right|\psi_{n}\right\rangle\right|^{2}}{E_{n}-E_{0}}\ ,
\label{lambda-matrix}
\end{equation}
where the sum runs over all $n$ orbital excited states \cite{Pryce50, Dai08}. As such, the energy difference in the denominator directly relates to the crystal field splitting of the orbitals. It should be noted that if the two atoms in a dimer have perpendicular anisotropies (\textit{i.e.} in the $\left\{\frac{3}{2},\frac{1}{2}\right\}$ dimer), the assignment of the axes $x, y \ \mbox{and}\ z$ differ between $\hat{\mathcal{H}}^{\rm (A)}$ and $\hat{\mathcal{H}}^{\rm (B)}$ in accordance with the anisotropy axes.

Using the total Heisenberg Hamiltonian (\ref{Heisenberg-Hamiltonian}) we generated simulated IETS spectra as described in Refs. \cite{Hirjibehedin07} and \cite{Gauyacq12}, for all three Fe dimers. As discussed below, $\Lambda_{zz}$ is much more sensitive to local strain than $\Lambda_{xx}$ and $\Lambda_{yy}$. Therefore, the only parameters that were allowed to vary between atoms were $\Lambda_{zz}$ and $J$. The transverse anisotropy parameter $E$ was not changed from its isolated-atom value $0.31$ meV \cite{Hirjibehedin07}. By choosing parameter values as summarized in Table~\ref{table1} we find excellent agreement with the measured spectra on each atom for all magnetic field directions, as shown in Figs.~1d--f and Fig.~S1.

\begin{table}[htbp]
%\begin{center}
\centering\tabcolsep=1.2mm
\begin{tabular}{*5{c}}
\ & \includegraphics[width=1.8cm]{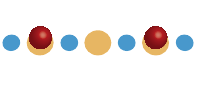} & \multicolumn{2}{c}{\includegraphics[width=1.8cm]{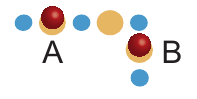}} & \includegraphics[width=1.8cm]{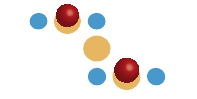} \\
\vspace{0.3cm}
\textbf{Dimer type} & \textbf{\{2,0\}} & \multicolumn{2}{c}{
${\bf {\left\{\frac{3}{2},\frac{1}{2}\right\}}}$
} & \textbf{\{1,1\}} \\
\hline
%\vspace{0.2cm}
\ & \ & A & B & \ \\
$J$ (meV) & $+0.70$ & \multicolumn{2}{c}{$+0.20$} & $-0.69$\\
$\Lambda_{zz}$ (eV$^{-1}$) & $14.2$ & $13.1$ & $10.5$ & $10.9$\\
$D$ (meV) & $-1.87$ & $-1.70$ & $-1.30$ & $-1.37$\\
$\Delta D$ (meV) & $-0.32$ & $-0.15$ & $+0.25$ & $+0.18$
\end{tabular}
%\end{center}
\caption{Values for $J$, $\Lambda_{zz}$ and $D$ (which follows from $\Lambda_{zz}$) found for each atom in all three types of Fe dimer, studied by fitting measured IETS spectra with simulated spectra. Only in the case of the $\left\{\frac{3}{2},\frac{1}{2}\right\}$ dimer are atoms A and B distinguishable. In the last row, for each atom the anisotropy energy shift $\Delta D$ with respect to the isolated atom value $D =-1.55$ meV is presented. All values shown are based on $\Lambda_{xx} = 0$ and $\Lambda_{yy} = 4.0$ eV$^{-1}$. Due to a linear dependence between $\Lambda_{xx}$, $\Lambda_{yy}$ and $\Lambda_{zz}$, practically identical simulated spectra can be obtained by adding a constant up to $2$ eV$^{-1}$ to each $\Lambda_{\mu\mu}$. This constant has no influence on the values for $J$ and~$D$.}
\label{table1}
\end{table}

Close inspection of the obtained parameter values reveals that the magnitude of the uniaxial anisotropy parameter $|D|$ has increased compared to its isolated-atom value $D=-1.55$ meV \cite{Hirjibehedin07} for both atoms in the $\left\{2,0\right\}$ dimer and for atom A in the $\left\{\frac{3}{2},\frac{1}{2}\right\}$ dimer. In contrast, it has decreased for atom B in the $\left\{\frac{3}{2},\frac{1}{2}\right\}$ dimer and for both atoms in the $\left\{1,1\right\}$ dimer. It appears therefore, that the magnitude of an atom's uniaxial anisotropy increases when a second atom is placed along (or within one bond length of) its easy axis, and that it decreases for those atoms where the second atom is further away from the easy axis.

In order to explain the observed behavior qualitatively we consider the immediate environment of the Fe atom, specifically its two nearest-neighbor N atoms, as shown in Fig.~2a. According to density functional theory (DFT) calculations, in the case of an isolated Fe atom on Cu$_2$N the N atoms are situated slightly lower than the Fe atom \footnote{The calculated N--Fe--N angle varies between models from $~130^{\circ}$ \cite{Hirjibehedin07} to $~175^{\circ}$ \cite{Nicklas11}.}. The resulting crystal field has $C_{2v}$ symmetry, which breaks all degeneracies between the five \textit{d}-orbitals \cite{Cotton90} (Fig.~2b). In the perturbative limit (\textit{i.e.} if the crystal field splitting is large compared to the spin-orbit coupling), the unquenched orbital momentum along the $z$-axis, $\Lambda_{zz}$, is inversely proportional to the energy difference $\Delta E_1$ between the $d_{xy}$ and $d_{x^2-y^2}$ orbitals. This means that if the ${\rm N}-{\rm Fe}-{\rm N}$ angle becomes closer to $180^{\circ}$, $\Lambda_{zz}$ and therefore $|D|$ rapidly increases. Likewise, if the angle becomes smaller, $|D|$ decreases. In contrast, $\Lambda_{xx}$ and $\Lambda_{yy}$ relate to the energy differences $\Delta E_2$ and $\Delta E_3$ (Fig.~2b), which will depend much less dramatically on the bond angle.

\begin{figure}[b]
\begin{centering}
\includegraphics{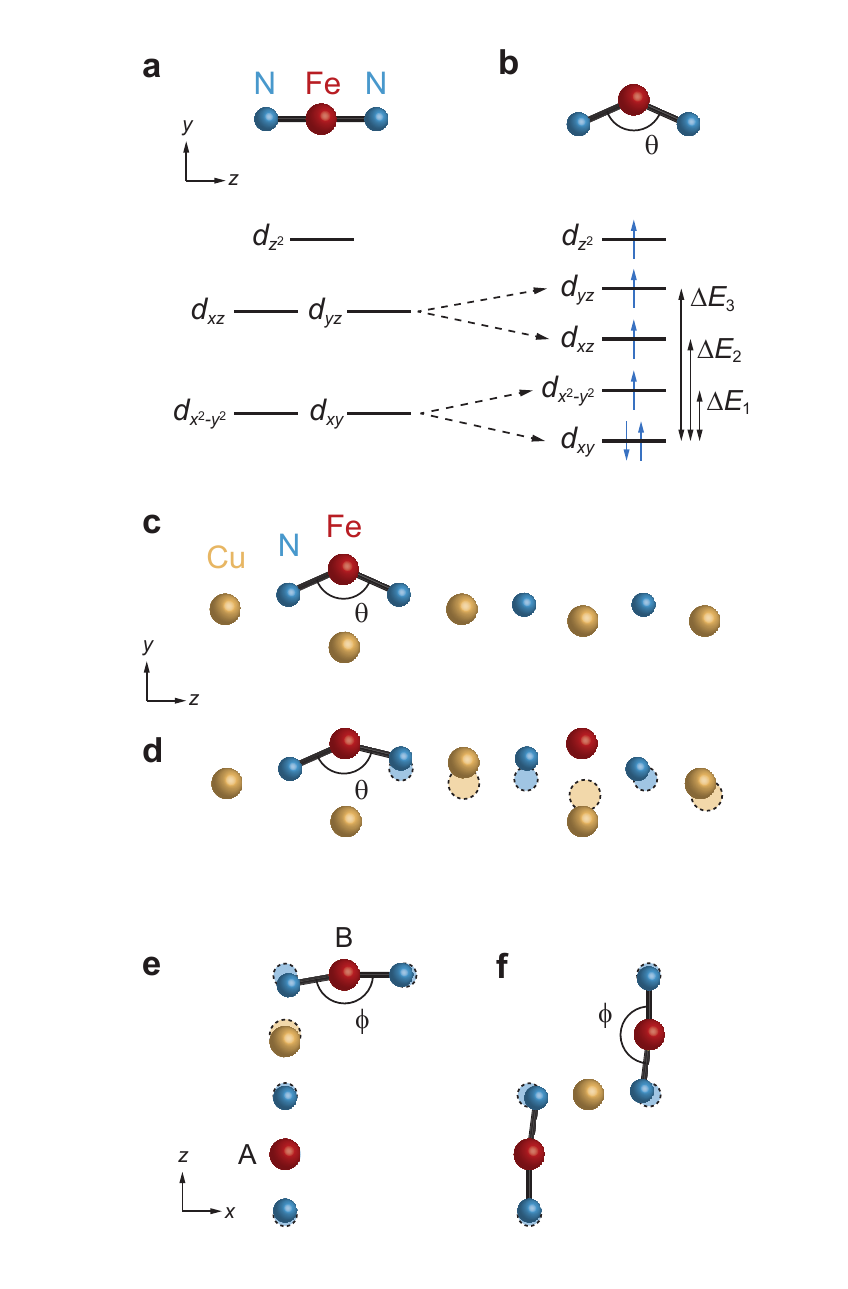}
\caption{(color online). (a) Splitting of d orbital energies in a linear crystal field. (b) Splitting of d orbital energies in a crystal field where the ligand-metal-ligand angle $\theta$ is slightly smaller than $180^{\circ}$. (c) Schematic of a single Fe adsorbed onto Cu$_2$N, after structure calculated using DFT \cite{Hirjibehedin07}: a section in the $yz$ plane along the \mbox{Cu--N} direction is shown. (d) Equivalent schematic section of $\left\{2,0\right\}$ Fe dimer: dashed outlines show atom positions for the single Fe case. Since the central Cu atom is lifted due to bonding to two Fe atoms, $\theta$ is larger (closer to $180^{\circ}$) for the dimer than for a single Fe atom. (e) Schematic of a $\left\{\frac{3}{2},\frac{1}{2}\right\}$ Fe dimer in the $xz$ plane. For a single Fe atom adsorbed onto Cu$_2$N the \mbox{N--Fe--N} angle in the $xz$ plane $\phi=180^{\circ}$: in the case of the $\left\{\frac{3}{2},\frac{1}{2}\right\}$ dimer the presence of Fe A compresses bond lengths along the \mbox{Cu--N} direction (as in (d)), so that for Fe B $\phi<180^{\circ}$. Similarly, for the $\left\{1,1\right\}$ dimer (f), for both Fe atoms $\phi<180^{\circ}$. In (e) and (f) dashed outlines show the unperturbed Cu$_2$N lattice. Atomic displacements are only indicative and are not to scale.}
\label{Figure2}
\end{centering}
\end{figure}

\begin{figure}[htbp]
\begin{centering}
\includegraphics{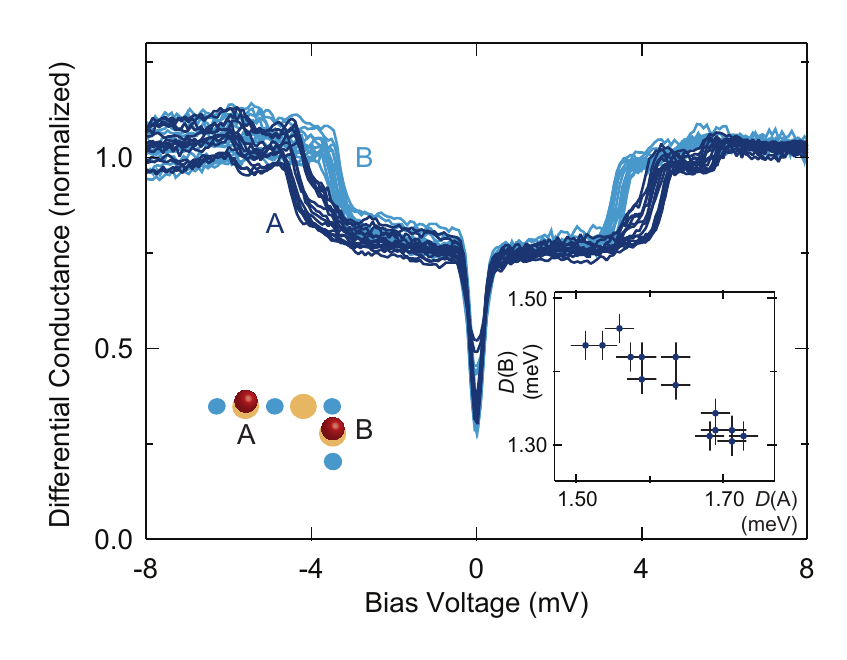}
\caption{(color online). IETS measurements performed on 14 different instances of the type $\left\{\frac{3}{2},\frac{1}{2}\right\}$ dimer at zero magnetic field, showing a spread of the step positions for both atoms in the dimer.  In the inset, the anisotropies of the two atoms of each dimer, $D^{\rm (A)}$ and $D^{\rm (B)}$, obtained by fitting the measured IETS spectra, are plotted against each other, indicating a strong anti-correlation between the two atoms (with Pearson correlation coefficient $R=-0.93$). The error bars represent the fitting uncertainty for $D$.}
\label{Figure3}
\end{centering}
\end{figure}

Figs.~2c--f demonstrate how the observed variations in $D$ in our dimers could be explained in terms of this \mbox{N--Fe--N} angle. When an Fe atom is deposited on Cu$_2$N, the neighboring N atoms and the next-nearest neighbor Cu atoms are pulled upward and toward the Fe atom \cite{Hirjibehedin07}. In the case of the $\left\{2,0\right\}$ dimer, we may expect that the central Cu atom will be raised even more, due to bonding to two Fe atoms, leading to an increased \mbox{N--Fe--N} angle (Fig.~2d) and hence increased $|D|$. The same reasoning applies to atom A of the $\left\{\frac{3}{2},\frac{1}{2}\right\}$ dimer. For atom B of this dimer the situation is different. In this case we expect the N atom to be pulled sideways due to the strain caused by atom A, as a result of which the \mbox{N--Fe--N} angle in the horizontal plane decreases (Fig.~2e), leading to a decreased $|D|$ value. A similar situation occurs for both atoms of the $\left\{1,1\right\}$ dimer (Fig.~2f).

It is possible to estimate the orbital energy splittings based on the measured values of $D$, $E$ and $g_{\mu}$. Unfortunately, all three values of $\Lambda_{\mu\mu}$ can only be fully constrained using measurements at very high magnetic fields. In the current situation we are left with a linear dependence between $\Lambda_{xx}$, $\Lambda_{yy}$ and $\Lambda_{zz}$. Based on measurements taken on an isolated Fe atom in magnetic fields up to 7 T in all three directions \cite{Hirjibehedin07}, we estimate that $\Lambda_{xx} < 2$~eV$^{-1}$. Using this value, we find that $\Delta E_1 = 305 \pm 25$ meV for the isolated Fe atom and ranges from $260 \pm 25$ meV for the atoms in the $\left\{2,0\right\}$ dimer to $350 \pm 30$ meV for atom B of the $\left\{\frac{3}{2},\frac{1}{2}\right\}$ dimer. The magnitude of these values compared to the used value for the spin-orbit constant $\lambda = -12.4$ meV \cite{Dai08} justifies our choice to treat spin-orbit coupling as a perturbation with respect to the crystal field.

It was observed that the $\left\{\frac{3}{2},\frac{1}{2}\right\}$ dimer shows some variation in anisotropy values from one instance of the dimer to another. Fig.~3 shows IETS spectra taken on both atoms of $14$ different instances of this dimer at zero magnetic field, revealing a spread of step positions. By performing the same fitting procedure as for the spectra in Fig.~1, we find $D$ values varying from the mean by $\pm 6 \%$. The anisotropy shifts are anti-correlated between atoms A and B (see Fig.~3, inset): dimers showing larger shifts of $D^{\rm (A)}$ away from the isolated-atom value $D=-1.55$ meV also have larger shifts of $D^{\rm (B)}$ in the opposite direction. This anti-correlation suggests a variation in local \emph{lattice} strain, applied to the whole dimer: if the local lattice is expanded, the strain induced by each Fe atom in the dimer on the other is reduced, and both Fe atoms' $D$ values  move closer to the isolated-atom value. The range of anisotropy values is similar to that found for single Fe atoms, which was attributed to the variation in lattice strain across the Cu$_2$N islands \cite{Hirjibehedin07,Ohno03}. However, we found no correlation between the variation in anisotropy and the position or orientation of each dimer on the Cu$_2$N island, nor with the size of the Cu$_2$N islands. Possibly, subsurface defects may play a role in the variation of the local environment of the Fe dimers.

In the current study the simulated IETS spectra were produced using a purely isotropic Heisenberg coupling parameter $J$. This is in contrast with previous results for the $\left\{2,0\right\}$ dimer, in which an Ising coupling parameter $J_z=+1.2$~meV was found \cite{Loth12}. In conjunction with increased values for $|D|$, we find a lower value $J=+0.70$~meV for this dimer. Our finding of strain-enhanced $D$ for Fe dimers is supported by the observation of increased spin excitation energies and spin lifetimes for a dimer composed of Fe and a non-magnetic atom \cite{Loth10}. 

In addition to the observed changes in $D$, Table~\ref{table1} also indicates variation in the strength and sign of the coupling $J$, which changes from antiferromagnetic in the $\left\{2,0\right\}$ and $\left\{\frac{3}{2},\frac{1}{2}\right\}$ dimers to ferromagnetic in the $\left\{1,1\right\}$ dimer. The physical origin for $J$ is likely to be a combination of superexchange and RKKY, as explored via DFT modeling for Gd atoms on Cu$_2$N \cite{Lin12}. The ability to tune the sign of the interatomic magnetic coupling through atom manipulation will enable atomically engineered spin structures to be designed in which a broad range of magnetic phenomena are realized. 

In summary, we have demonstrated that the magneto-crystalline anisotropy of a magnetic adatom embedded in a covalent surface network can be controllably enhanced or reduced by positioning a second adatom nearby. Strain due to the presence of this second adatom will cause a slight change in the angle between the magnetic atom and its neighboring ligands. Using a qualitative first principles model we demonstrate that the magnetic anisotropy depends highly sensitively on this angle. We suggest that this strain-enhanced anisotropy may play a critical role in the reported magnetic stabilization of atomically-assembled antiferromagnetic structures \cite{Loth12}, which is essential to the creation of atomic-scale magnetic memory devices.  

We thank J.~Fern\'andez-Rossier, C.~F.~Hirjibehedin, A.~J.~Heinrich and G.~A.~Steele for discussions and R. Hoogerheide for technical support. This work was supported by the Dutch Organization for Fundamental Research on Matter (FOM), and by the Kavli Foundation.

\bibliography{library}

\end{document}